\documentclass[letterpaper,showpacs,preprintnumbers,eqsecnum,superscriptaddress,aps,nofootinbib]{revtex4}
\usepackage{amssymb}\usepackage[centertags]{amsmath}\usepackage{txfonts}\usepackage{epsfig}\usepackage{bm}
\usepackage{color}\usepackage{graphicx,graphics}\usepackage{multirow}\usepackage{float}\usepackage{ulem}
\usepackage[dvipdfm,pdfstartview=FitH]{hyperref}\usepackage{setspace}\usepackage{slashed}\usepackage{booktabs}\usepackage{CJK}
\hypersetup{colorlinks=true, citecolor=blue, linkcolor=red, filecolor=black,urlcolor=blue}
\allowdisplaybreaks[4]

\begin{document}

\title{Spin Density Matrix for Spin-$\frac{3}{2}$ Particles}

\author{Jiao Jiao Song}

\affiliation{School of Physics, Shandong University, Jinan, Shandong 250100, China}
\affiliation{Institute of High Energy Physics, Beijing 100049, China}

\author{Wan Ling Chang}
\affiliation{Institute of High Energy Physics, Beijing 100049, China}

\begin{abstract}
  The description of particles with spin can be attained by using a spin density matrix in high energy reaction. In this paper we present a parametrization of the spin density matrix for spin-3/2 particles in the Cartesian form. Comparing the standard form which is given by the spherical tensor operator with the Cartesian form yields four equations for the spin polarizations. These four equations can be used to eliminate the dependent spin polarizations. We also present the probabilistic interpretations of these spin polarizations by defining the spin operator in the spherical coordinate frame. These results can be used in the cross section to calculate the hadron polarizations. To illustrate this, we calculate the inclusive electron-positron annihilation process in terms of spin polarizations.
  \\
  \\
  $Keywords$: spin density matrix, Cartesian form, probabilistic interpretation
\end{abstract}

\pacs{13.66.Bc, 13.85.Ni, 13.87.Fh, 21.10.Hw}

\maketitle

\section{Introduction}

Since the establishment of the quantum chromodynamics, physicists have been striving for fully understanding of the hadron structures and the nonperturbative hadronization processes. However, we do not know them very well, e.g, the proton size \cite{Carlson:2015jba}, the spin and the intrinsic structure of nucleons \cite{Aidala:2012mv,Deur:2018roz,Miller:2007uy}. Thanks to the factorization theorem \cite{Collins:1989gx}, many high energy reactions, deeply inelastic scattering, Drell-Yan and electron-positron annihilation processes, can be calculated in terms of the parton distribution and fragmentation functions. Particles in these reactions always have nonzero spins. The description of particles with spin can be attained by using a spin density matrix (SDM) in the rest frame of the particle, and the expansion coefficients represent the polarization of itself \cite{Bourrely:1980mr}.

In quantum mechanics, a series of problems can be resolved after the state vectors are settled down. Furthermore, density matrix, which can be applied in many areas, such as quantum physics, statistical physics, extends the concept of state. If the system under considering includes particles with spin, the description of these particles can be attained by using the so-called $spin~ density ~matrix$. It is a $(2s+1)\times(2s+1)$ matrix, where s is the particle spin \cite{Leader:2001gr}. According to the definition of the matrix, one can find out that $\rho$ is a hermitian matrix whose trace is unity. As a result of the hermitian property, there is a unitary matrix $U$ that can diagonalize it, i.e., $\rho^D=U^-1\rho U$, where $\rho^D$ denotes the diagonal matrix. For the density matrix, one can also prove that the diagonal elements are positive semi-definite and $\mathrm{Tr} \rho^2 \leq 1$.
The SDM can be written as a standard form by using the spherical tensor operator or a Cartesian form by using the irreducible spin tensor. For example, the best known density matrix of spin-1/2 particles can be decomposed on a Cartesian basis as
\begin{align}
  \rho = \frac{1}{2}(1+ \mathbf{P}\cdot \mathbf{\sigma}), \label{f:rhocartesin}
\end{align}
where $\mathbf{\sigma}$ is the Pauli matrix, $\mathbf{P}$ is the spin-polarization vector for the ensemble. The spin-1 density matrix can be parametrized in a similar way with the identity matrix, three spin vector operator matrices, $\Sigma^{x,y,z}$, and five spin tensors, $\Sigma^{ij}$ \cite{Bacchetta:2000jk}. In this paper, we focus on the parametrization of the density matrix of the spin-3/2 particles. We construct the density matrix  by using the spin operator in cartesian coordinate system. We also present the average value of the spin operator which can be expressed by the probabilities of some spin state.

Recently, Perotti, Faldt, Kupsc, Leupold and Song discussed the polarization observables for baryon-antibaryon pair productions in electron-positron annihilation process with the density matrix \cite{Perotti:2018wxm}. The helicity formalism used in \cite{Perotti:2018wxm} was originally developed by Jacob and Wick for hadron productions \cite{Jacob:1959at}. Perotti and colleagues discussed the density matrix from the point of view of experimental measurement. They mainly provided modular tools to construct joint decay distributions of sequential decay processes for the spin-1/2 and spin-3/2 baryon pair productions in electron-positron annihilation process. However, in this paper we consider the parametrization of the SDM for inclusive spin-3/2 particles in Cartesian form. Our consideration emphasizes on the quark fragmentation or hadronization process. The SDM considered here can not only be applied to the annihilation process but also applied to any other spin-3/2 baryon production processes. To be explicit, our consideration applies to the direct produced baryons, e.g, $\Omega$. Hyperon $\Omega$ which is spin-3/2 is made up of three strange quarks. Measuring the polarization of hyperon in the high energy reactions is an ideal way to study the spin transfer in hadronization processes \cite{Adam:2018kzl,Adam:2018wce}, which can help us to understand more about the hadronization.

The rest of the paper is organized as follows. We first calculate the parametrization of the spin-3/2 SDM in Cartesian form in sect. \ref{S:sdm}. In the following section, we present probabilistic interpretations of these spin polarizations. The brief introduction to the application of the SDM in high energy reactions will be given in sect. \ref{S:application}. A brief summary is given in sect. \ref{S:summary}.

\section{Spin Density Matrix in Cartesian Form}\label{S:sdm}

As mentioned in the introduction, the SDM can either be written as the standard form in terms of the spherical tensor operator or be given in a Cartesian form.
For the standard form, using the spherical tensor, the SDM can be expended as
\begin{align}
  &\rho=\frac{1}{2s+1}\sum_{L,M}(2L+1){t^L_M}^*T^L_M, \label{f:rhostandrd}
\end{align}
where $T^L_M$ is the spherical tensor operator, L is the rank of the tensor operator which satisfies $0\leq L \leq 2s$, $-L\leq M \leq L$, $t^L_M$ is known as multipole parameters.
In this paper, we only consider the SDM for spin-3/2 particles in the Cartesian form. It is shown that the SDM can be defined by the identity matrix, three spin matrices $\Sigma^i$ (generalization of the Pauli matrices to the four-dimensional case) and spin tensors. Since the SDM is a $4 \times 4$ matrix, it can be parameterized as
\begin{align}
  \rho=a1+bS^i\Sigma^i+cS^{ij}\Sigma^{ij}+dS^{ijk}\Sigma^{ijk}, \label{f:rhomatrix}
\end{align}
where $1$ denotes identity matrix and $\Sigma^i$ are spin operators,
\begin{align}
  &\Sigma^x=\frac{1}{2}\left(
        \begin{array}{cccc}
          0 & \sqrt{3} & 0 & 0 \\
          \sqrt{3} & 0 & 2 & 0 \\
          0 & 2 & 0 & \sqrt{3} \\
          0 & 0 & \sqrt{3} & 0 \\
        \end{array}
      \right),
 \quad  \Sigma^y=\frac{i}{2}\left(
        \begin{array}{cccc}
          0 & -\sqrt{3} & 0 & 0 \\
          \sqrt{3} & 0 & -2 & 0 \\
          0 & 2 & 0 & -\sqrt{3} \\
          0 & 0 & \sqrt{3} & 0 \\
        \end{array}
      \right),
 \quad \Sigma^z=\frac{1}{2}\left(
                   \begin{array}{cccc}
                     3 & 0 & 0 & 0 \\
                     0 & 1 & 0 & 0 \\
                     0 & 0 & -1 & 0 \\
                     0 & 0 & 0 & -3 \\
                   \end{array}
                 \right). \label{f:spinvector}
\end{align}
The second order tensor $\Sigma^{ij}$ is defined as
\begin{align}
  &\Sigma^{ij}=\frac{1}{2}(\Sigma^i\Sigma^j+\Sigma^j\Sigma^i)-\frac{5}{4} \delta^{ij}. \label{f:sigmaij}
\end{align}
Here $\Sigma^{ij}$ satisfy $\Sigma^{xx}+\Sigma^{yy}+\Sigma^{zz}=0$. For $\Sigma^{ijk}$, it is defined as
\begin{align}
  \Sigma^{ijk}=\frac{1}{6}\big[&(\Sigma^i\Sigma^j\Sigma^k+\Sigma^i\Sigma^k\Sigma^j
  +\Sigma^j\Sigma^i\Sigma^k+\Sigma^j\Sigma^k\Sigma^i+\Sigma^k\Sigma^i\Sigma^j\nonumber\\
  +&\Sigma^k\Sigma^j\Sigma^i)-\frac{5}{12}(\delta^{ij}\Sigma^k+\delta^{jk}\Sigma^i
  +\delta^{ki}\Sigma^j)\big]. \label{f:sigmaijk}
\end{align}
In Eq. (\ref{f:rhomatrix}), $S^i, S^{ij}$ and $S^{ijk}$ denote the corresponding spin-polarization vectors and tensors. The coefficients $a, b, c$ and $d$ are determined by the following equations,
\begin{align}
  &Tr[\rho]=1, \\
  &Tr[\rho\Sigma^i]=5S^i, \\
  &Tr[\rho\Sigma^{ij}]=S^{ij}, \\
  &Tr[\rho\Sigma^{ijk}]=\frac{1}{4}S^{ijk}.
\end{align}
According to the above formulae, we obtain $a=\frac{1}{4}, b=1, c=\frac{1}{3}, d=\frac{1}{3}$.
As a result, the density matrix can be rewritten as
\begin{align}
  \rho=\frac{1}{4}\big[1+4S^i\Sigma^i+\frac{4}{3}S^{ij}\Sigma^{ij} +\frac{4}{3}S^{ijk}\Sigma^{ijk}\big]. \label{f:rho3/2}
\end{align}

The definition of $\Sigma^{ij}$ in Eq. (\ref{f:sigmaij}) yields 9 components. However, due to the exchange symmetry constraint $(i \leftrightarrow j)$, only 6 of them are left. By using $\Sigma^{xx}+\Sigma^{yy}+\Sigma^{zz}=0 $, one obtains 5 components in total, $\Sigma^{xx},\Sigma^{xy},\Sigma^{xz},\Sigma^{yz},\Sigma^{zz}$, which correspond to 5 coefficients. For $\Sigma^{ijk}$, by using the exchange symmetry constraint and the following equations
\begin{align}
  &\Sigma^{xxx}+\Sigma^{xyy}+\Sigma^{xzz}=\frac{4}{3}\Sigma^x,\\
  &\Sigma^{yxx}+\Sigma^{yyy}+\Sigma^{yzz}=\frac{4}{3}\Sigma^y,\\
  &\Sigma^{zxx}+\Sigma^{zyy}+\Sigma^{zzz}=\frac{4}{3}\Sigma^z,
\end{align}
one obtains 7 components in total, $\Sigma^{xxx}$, $\Sigma^{xxy}$, $\Sigma^{yzz}$, $\Sigma^{zzz}$, $\Sigma^{zxx}$, $\Sigma^{zzx}$, $\Sigma^{xyz}$, which correspond to 7 coefficients. One can choose other combinations of these components as long as $\Sigma^{xyz}$ is included. From the above discussion, we obtain 16 tensor matrices including the unit one. In this case, there are 16 coefficients which correspond to these components. We show them in the following context.

We introduce the spherical tensor operators.
\begin{align}
  &T^0_0=1, \label{f:T00} \\
  &T^1_0=c^1_0\Sigma^z,\\
  &T^1_{\pm1}=\mp c^1_{1}(\Sigma^x\pm i\Sigma^y),\\
  &T^2_0=c^2_0 (2\Sigma^{zz}-\Sigma^{xx}-\Sigma^{yy}),\\
  &T^2_{\pm1}=\mp c^2_{1}(\Sigma^{xz}\pm i\Sigma^{yz}),\\
  &T^2_{\pm2}=c^2_{2}(\Sigma^{xx}-\Sigma^{yy}\pm 2i\Sigma^{xy}),\\
  &T^3_0=c^3_0(2\Sigma^{zzz}-3\Sigma^{zxx}-3\Sigma^{zyy}),\\
  &T^3_{\pm1}=\mp c^3_1\Big[(4\Sigma^{xzz}-\Sigma^{xxx}-\Sigma^{xyy})\pm i(4\Sigma^{yzz}-\Sigma^{yxx}-\Sigma^{yyy})\Big],\\
  &T^3_{\pm2}=c^3_2(\Sigma^{xxz}-\Sigma^{yyz}\pm 2i\Sigma^{xyz}),\\
  &T^3_{\pm3}=\mp c^3_3(\Sigma^{xxx}\pm3i\Sigma^{xxy}-3\Sigma^{xyy}\mp i\Sigma^{yyy}). \label{f:T33}
\end{align}
Using Eqs. (\ref{f:T00})-(\ref{f:T33}), Eqs. (\ref{f:rhostandrd}) and (\ref{f:rho3/2}), the following equations can be obtained.
\begin{align}
  &t^0_0=1 , \label{f:t00} \\
  &S_L=\frac{15c^1_0}{4}t^1_0,\\
  &S_T^x=-\frac{15c^1_1}{4}(t^1_1-t^1_{-1}),\\
  &S_T^y=-i\frac{15c^1_1}{4}(t^1_{-1}+t^1_1),\\
  &S_{LL}=\frac{15c^2_0}{2}t^2_0;,\\
  &S_{LT}^{x}=-\frac{15c^2_1}{4}(t^2_{1}-t^2_{-1}),\\
  &S_{LT}^{y}=-i\frac{15c^2_1}{4}(t^2_{-1}+t^2_1),\\
  &S_{TT}^{xy}=i\frac{15c^2_2}{2}(t^2_2-t^2_{-2}),\\
  &S_{TT}^{xx}=\frac{15c^2_2}{4}(t^2_2+t^2_{-2})-\frac{15c^2_0}{4}t^2_0,\\
  &S_{TT}^{yy}=-\frac{15c^2_2}{4}(t^2_2+t^2_{-2})-\frac{15c^2_0}{4}t^2_0,\\
  &S_{LLL}=\frac{21c^3_0}{8}t^3_0,\\
  &S_{LLT}^x=-\frac{21c^3_1}{4}(t^3_{1}-t^3_{-1}),\\
  &S_{LLT}^y=-i\frac{21c^3_1}{4}(t^3_{1}+t^3_{1}),\\
  &S_{LTT}^{xy}=i\frac{21c^3_2}{8}(t^3_2-t^3_{-2}),\\
  &S_{LTT}^{xx}=\frac{21}{16}\left[c^3_2(t^3_2+t^3_{-2})-3c^3_0t^3_0\right],\\
  &S_{LTT}^{yy}=\frac{21}{16}\left[-c^3_2(t^3_2+t^3_{-2})-3c^3_0t^3_0\right],\\
  &S_{TTT}^{xxx}=\frac{21}{16}\left[c^3_1(t^3_{1}-t^3_{-1})-c^3_3(t^3_{3}-t^3_{-3})\right],\\
  &S_{TTT}^{xyy}=\frac{21}{16}\left[c^3_1(t^3_{1}-t^3_{-1})+3c^3_3(t^3_{3}-t^3_{-3})\right],\\
  &S_{TTT}^{yyy}=i\frac{21}{16}\left[c^3_1(t^3_{1}+t^3_{-1})+c^3_3(t^3_{3}+t^3_{-3})\right],\\
  &S_{TTT}^{yxx}=i\frac{21}{16}\left[c^3_1(t^3_{1}+t^3_{-1})-3c^3_3(t^3_{3}+t^3_{-3})\right].\label{f:t33}
\end{align}
These coefficients, $c^{1,2,3}_{0,\pm1,\pm2,\pm3}$, in Eqs. (\ref{f:t00})-(\ref{f:t33}) which can be normalized into the formal values do not affect the derivation, we won't show them in this paper.
Four equations are obtained,
\begin{align}
  &S_{LL}+S_{TT}^{xx}+S_{TT}^{yy}=0, \label{f:SLL}\\
  &3S_{LLL}+S_{LTT}^{xx}+S_{LTT}^{yy}=0, \\
  &3S_{TTT}^{xxx}+S_{TTT}^{xyy}+S_{LLT}^x=0,\\
  &3S_{TTT}^{yyy}+S_{TTT}^{yxx}+S_{LLT}^y=0, \label{f:STTT}
\end{align}
which can be used to eliminate 4 coefficients (polarizations). Finally, there are 16 kinds of polarizations obtained to describe the spin-3/2 density matrix. One unpolarized case, three spin vectors polarizations and the $5+7=12$ spin tensor polarizations.

Here the polarizations are named under the following conventions. Subscript $L, T$ are used to denote the longitudinal (z) and the transverse direction (x, y) components, respectively. Superscripts $x, y$ are used to distinguish the x- and y-components while $z$ is omitted.

We can show the explicit parametrization of the spin-3/2 SDM for that all the spin polarizations and the corresponding polarizations are known. Substituting all of the $S^{i,ij,ijk}$ and $\Sigma^{i,ij,ijk}$ into Eq. (\ref{f:rho3/2}) yields
\begin{align}
\rho=\frac{1}{4}\left(
\begin{array}{cccc}
  A_1 & B_1 & C_1 & D \\
  B_1^* & A_2 & B_2 & C_2 \\
  C_1^* & B_2^* & A_3 & B_3 \\
  D^* & C_2^* & B_3^* & A_4\\
\end{array}
\right), \label{f:rhoexpression}
\end{align}
where
\begin{align}
  &A_1=1+6S_L+\frac{4}{3}S_{LL}+S_{LLL}, \\
  &A_2=1+2S_L-\frac{4}{3}S_{LL}-3S_{LLL}, \\
  &A_3=1-2S_L-\frac{4}{3}S_{LL}+3S_{LLL}, \\
  &A_4=1-6S_L+\frac{4}{3}S_{LL}-S_{LLL},\\
  &B_1=2\sqrt{3}S_T^{x-iy}+ \frac{2\sqrt{3}}{3}S_{LT}^{x-iy}+\frac{\sqrt{3}}{3}S_{LLT}^{x-iy},\\
  &B_2=4S_T^{x-iy}-S_{LLT}^{x-iy},\\
  &B_3=2\sqrt{3}S_T^{x-iy}- \frac{2\sqrt{3}}{3}S_{LT}^{x-iy}+\frac{\sqrt{3}}{3}S_{LLT}^{x-iy},\\
  &C_1=\frac{2\sqrt{3}}{3}S_{TT}^{xx-ixy}+\frac{\sqrt{3}}{3}S_{LTT}^{xx-ixy}, \\
  &C_2=\frac{2\sqrt{3}}{3}S_{TT}^{xx-ixy}-\frac{\sqrt{3}}{3}S_{LTT}^{xx-ixy},\\
  &D=S_{TTT}^{xxx-ixxy}.
\end{align}
We note that $X^* (X=B, C, D)$ is the complex conjugate of $X$.
Here $S_{LT}^{x-iy}=S_{LT}^x-iS_{LT}^y$, the similar shorthanded notations are also used for $S_{TT}, S_{LLT}, S_{LTT}$ and $S_{TTT}$. In order to obtain the SDM, Eq. (\ref{f:rhoexpression}), we have used these relations shown in Eqs. (\ref{f:SLL})-(\ref{f:STTT}). Since the hermitian matrix whose diagonal elements are positive semi-definite, we can obtain the positivity bound of the longitudinal polarizations,
\begin{align}
  & -\frac{3}{5} \leq S_{L} \leq \frac{3}{5}, \\
  & -\frac{3}{4} \leq S_{LL} \leq \frac{3}{4}, \\
  & -\frac{3}{10} \leq S_{LLL} \leq \frac{3}{10}.
\end{align}
Here we define the total overall degree of polarization. The degree of polarization of any state which is proportional
to its distance to the unpolarized state is defined as \cite{Leader:2001gr,Doncel:1972ez}
\begin{align}
  d=\frac{1}{\sqrt{2s}}\sqrt{(2s+1)\mathrm{Tr}\rho^2-1}. \label{f:overall}
\end{align}
If only the longitudinal polarizations are taken into consideration, the overall degree of polarization is given by
\begin{align}
  d&=\frac{1}{3}\sqrt{60(S_{L})^2+\frac{16}{3}(S_{LL})^2+15(S_{LLL})^2}. \label{f:overalllongfull}
\end{align}
The complete the overall degree of polarization is
\begin{align}
  d&=\frac{1}{3}\bigg\{ 60\Big[(S_{T}^x)^2+(S_{y}^x)^2+(S_{L})^2\Big] \nonumber\\
  &+\frac{16}{3}(S_{LL})^2+4\Big[(S_{LT}^x)^2+(S_{LT}^y)^2+(S_{TT}^{xy})^2+4(S_{LTT}^{xy})^2\Big] \nonumber\\
  &+15(S_{LLL})^2+\frac{5}{2}(S_{LLT}^{x})^2+\frac{5}{2}(S_{LLT}^{y})^2+(S_{LTT}^{xx})^2+(S_{LTT}^{xy})^2 +\frac{3}{2}(S_{TTT}^{xxx})^2+\frac{3}{2}(S_{TTT}^{xxy})^2\bigg\}^{1/2}, \label{f:overallPolar}
\end{align}
whose value ranges between 0 and 1.


\section{Probabilistic interpretation} \label{S:probability}

In the previous section, we present the SDM in Cartesian form. It is parameterized by the spin vector (tensor) with the corresponding coefficients. For a particular component of the spin vector, it measures a combination of probabilities of finding the system in a certain spin state which is defined in the particle rest frame. For a system of hadrons, SDM is always being used to describe the polarization. From the original definition of the density matrix, it can be given by
\begin{align}
  \rho=\sum_{m,n}|n\rangle \langle n|~ \rho ~|m\rangle \langle m|=\sum_{m,n}\mathrm{Tr}\big[\rho |m\rangle \langle n|\big]|n\rangle \langle m|,\label{f:spindensity}
\end{align}
where $|m\rangle, |n\rangle$ are eigenstates of certain operators. If Eq. (\ref{f:spindensity}) is SDM, $|m\rangle, |n\rangle$ will be eigenstates of spin operators. The probability of finding one of these states is defined as $P\equiv\mathrm{Tr}\big[\rho |m\rangle \langle n|\big]$.

The spin operator can be defined in arbitrary direction,
\begin{align}
  \Sigma^i\hat n^i=\Sigma^x \sin\theta\cos\phi +\Sigma^y \sin\theta\sin\phi +\Sigma^z \cos\theta, \label{f:spinoperator}
\end{align}
where $\theta$ and $\phi$ denote the polar and azimuthal angles, respectively. The polarization of a system can be calculated with the following equation \cite{Yang:2019rrn},
\begin{align}
  O=\langle \mathcal{O} \rangle =\mathrm{Tr} [\mathcal{O}\rho], \label{f:spinO}
\end{align}
where $\mathcal{O}$ denotes any spin operator.

In order to show the explicit expression of these probabilistic interpretations of the spin polarization, we first decompose the tensors for spin-3/2 particles into the following forms,
\begin{align}
  \Sigma^{xyz}&=\frac{1}{6}\Big[(\Sigma^x+\Sigma^y+\Sigma^z)^3-(\Sigma^x+\Sigma^y)^3 -(\Sigma^x+\Sigma^z)^3-(\Sigma^y+\Sigma^z)^3+(\Sigma^x)^3+(\Sigma^y)^3+(\Sigma^z)^3\Big], \label{f:sigmaSxyz}\\
  \Sigma^{xxy}&=\frac{1}{3}\Big[(2\Sigma^x+\Sigma^y)^3 -2(\Sigma^x+\Sigma^y)^3-(2\Sigma^x)^3 +2(\Sigma^x)^3+(\Sigma^y)^3\Big]-\frac{5}{12}\Sigma^y,\\
  \Sigma^{yzz}&=\frac{1}{3}\Big[(2\Sigma^z+\Sigma^y)^3 -2(\Sigma^z+\Sigma^y)^3-(2\Sigma^z)^3 +2(\Sigma^z)^3+(\Sigma^y)^3\Big]-\frac{5}{12}\Sigma^y,\\
  \Sigma^{zzx}&=\frac{1}{3}\Big[(2\Sigma^z+\Sigma^x)^3 -2(\Sigma^z+\Sigma^x)^3-(2\Sigma^z)^3 +2(\Sigma^z)^3+(\Sigma^x)^3\Big]-\frac{5}{12}\Sigma^x,\\
  \Sigma^{zxx}&=\frac{1}{3}\Big[(2\Sigma^x+\Sigma^z)^3 -2(\Sigma^x+\Sigma^z)^3-(2\Sigma^x)^3 +2(\Sigma^x)^3+(\Sigma^z)^3\Big]-\frac{5}{12}\Sigma^z,\\
  \Sigma^{xxx}&=(\Sigma^x)^3-\frac{5}{4}\Sigma^x,\\
  \Sigma^{zzz}&=(\Sigma^z)^3-\frac{5}{4}\Sigma^z. \label{f:sigmaSzzz}
\end{align}

By using Eq. (\ref{f:rhoexpression}), we take $(\Sigma^x+\Sigma^y+\Sigma^z)^3$ for example to show the probabilistic interpretation. Assuming that $|\ m\rangle $ is the eigenstate of the spin operator, therefore
\begin{align}
  \frac{1}{\sqrt{3}}(\Sigma^x+\Sigma^y+\Sigma^z)|\ m\rangle =m |\ m\rangle
\end{align}
where $m=\pm\frac{3}{2}, \pm\frac{1}{2}$ for spin-3/2 particles. Let's consider the expectation value of $(\Sigma^x+\Sigma^y+\Sigma^z)^3 $. Here we define $S_{LTT}^{xy}=\langle \Sigma^{xyz}\rangle$. For the other polarizations, the definitions are the same, see Eq. (\ref{f:spinO}).
\begin{align}
  &\ \langle(\Sigma^x+\Sigma^y+\Sigma^z)^3\rangle={\rm Tr}\big[\rho(\Sigma^x+\Sigma^y+\Sigma^z)^3\big]\nonumber\\
  &={\rm Tr}\bigg[\sum_{m,n}{\rm Tr}\big[\rho|\ n\rangle\langle m\ |\big]|\ m\rangle\langle n\ |(\Sigma^x+\Sigma^y+\Sigma^z)^3\bigg]\nonumber\\
  &=\sum_{m,n}{\rm Tr}\big[\rho|\ n\rangle\langle m\ |\big]{\rm Tr}\big[(\Sigma^x+\Sigma^y+\Sigma^z)^3|\ m\rangle\langle n\ |\big]\nonumber\\
  &=(\sqrt{3}m)^3\sum_{m,n}{\rm Tr}\big[\rho|\ n\rangle\langle m\ |\big]\delta_{mn}\nonumber\\
  &=\bigg(\frac{3\sqrt{3}}{2}\bigg)^3[P_{\frac{3}{2}}(\theta,\phi)-P_{-\frac{3}{2}}(\theta,\phi)]   +\bigg(\frac{\sqrt{3}}{2}\bigg)^3[P_{\frac{1}{2}}(\theta,\phi)-P_{-\frac{1}{2}}(\theta,\phi)], \label{f:sigmaxyz3}
\end{align}
where $(\theta,\phi)=(\arctan\sqrt{2},\frac{\pi}{4})$. Following the same steps, we can calculate all of the expectation values of these spin tensor. For $\Sigma^{xyz}, \Sigma^{xxy}$, we have

 \begin{align}
   \langle\Sigma^{xyz}\rangle &=\frac{\sqrt{3}}{2}\bigg(\frac{3}{2}\bigg)^3[P_{\frac{3}{2}}(\theta,\phi) -P_{-\frac{3}{2}}(\theta,\phi)] +\frac{\sqrt{3}}{2}\bigg(\frac{1}{2}\bigg)^3[P_{\frac{1}{2}}(\theta,\phi) -P_{-\frac{1}{2}}(\theta,\phi)] \nonumber\\
   &-\frac{\sqrt{2}}{3}\bigg(\frac{3}{2}\bigg)^3[P_{\frac{3}{2}}(\frac{\pi}{2},\frac{\pi}{4}) -P_{-\frac{3}{2}}(\frac{\pi}{2},\frac{\pi}{4})] -\frac{\sqrt{2}}{3}\bigg(\frac{1}{2}\bigg)^3[P_{\frac{1}{2}}(\frac{\pi}{2},\frac{\pi}{4}) -P_{-\frac{1}{2}}(\frac{\pi}{2},\frac{\pi}{4})] \nonumber\\
   &-\frac{\sqrt{2}}{3}\bigg(\frac{3}{2}\bigg)^3[P_{\frac{3}{2}}(\frac{\pi}{4},0) -P_{-\frac{3}{2}}(\frac{\pi}{4},0)] -\frac{\sqrt{2}}{3}\bigg(\frac{1}{2}\bigg)^3[P_{\frac{1}{2}}(\frac{\pi}{4},0) -P_{-\frac{1}{2}}(\frac{\pi}{4},0)] \nonumber\\
   &-\frac{\sqrt{2}}{3}\bigg(\frac{3}{2}\bigg)^3[P_{\frac{3}{2}}(\frac{\pi}{4},\frac{\pi}{2}) -P_{-\frac{3}{2}}(\frac{\pi}{4},\frac{\pi}{2})] -\frac{\sqrt{2}}{3}\bigg(\frac{1}{2}\bigg)^3[P_{\frac{1}{2}}(\frac{\pi}{4},\frac{\pi}{2}) -P_{-\frac{1}{2}}(\frac{\pi}{4},\frac{\pi}{2})] \nonumber\\
   &+\frac{1}{6}\bigg(\frac{3}{2}\bigg)^3[P_{\frac{3}{2}}(\frac{\pi}{2},0) -P_{-\frac{3}{2}}(\frac{\pi}{2},0)] +\frac{1}{6}\bigg(\frac{1}{2}\bigg)^3[P_{\frac{1}{2}}(\frac{\pi}{2},0) -P_{-\frac{1}{2}}(\frac{\pi}{2},0)] \nonumber\\
   &+\frac{1}{6}\bigg(\frac{3}{2}\bigg)^3[P_{\frac{3}{2}}(\frac{\pi}{2},\frac{\pi}{2}) -P_{-\frac{3}{2}}(\frac{\pi}{2},\frac{\pi}{2})] +\frac{1}{6}\bigg(\frac{1}{2}\bigg)^3[P_{\frac{1}{2}}(\frac{\pi}{2},\frac{\pi}{2}) -P_{-\frac{1}{2}}(\frac{\pi}{2},\frac{\pi}{2})] \nonumber\\
   &+\frac{1}{6}\bigg(\frac{3}{2}\bigg)^3[P_{\frac{3}{2}}(0,0) -P_{-\frac{3}{2}}(0,0)] +\frac{1}{6}\bigg(\frac{1}{2}\bigg)^3[P_{\frac{1}{2}}(0,0) -P_{-\frac{1}{2}}(0,0)], \label{f:meansigmaxyz}
 \end{align}
\begin{align}
   \langle\Sigma^{xxy}\rangle
   &=\frac{5\sqrt{5}}{3}\bigg(\frac{3}{2}\bigg)^3[P_{\frac{3}{2}}(\frac{\pi}{2},\phi_x) -P_{-\frac{3}{2}}(\frac{\pi}{2},\phi_x)] +\frac{5\sqrt{5}}{3}\bigg(\frac{1}{2}\bigg)^3[P_{\frac{1}{2}}(\frac{\pi}{2},\phi_x) -P_{-\frac{1}{2}}(\frac{\pi}{2},\phi_x)] \nonumber\\
   &-\frac{4\sqrt{2}}{3}\bigg(\frac{3}{2}\bigg)^3[P_{\frac{3}{2}}(\frac{\pi}{2},\frac{\pi}{4}) -P_{-\frac{3}{2}}(\frac{\pi}{2},\frac{\pi}{4})] -\frac{4\sqrt{2}}{3}\bigg(\frac{1}{2}\bigg)^3[P_{\frac{1}{2}}(\frac{\pi}{2},\frac{\pi}{4}) -P_{-\frac{1}{2}}(\frac{\pi}{2},\frac{\pi}{4})] \nonumber\\
   &-2\bigg(\frac{3}{2}\bigg)^3[P_{\frac{3}{2}}(\frac{\pi}{2},0) -P_{-\frac{3}{2}}(\frac{\pi}{2},0)] -2\bigg(\frac{1}{2}\bigg)^3[P_{\frac{1}{2}}(\frac{\pi}{2},0) -P_{-\frac{1}{2}}(\frac{\pi}{2},0)] \nonumber\\
   &+\frac{1}{3}\bigg(\frac{3}{2}\bigg)^3[P_{\frac{3}{2}}(\frac{\pi}{2},\frac{\pi}{2}) -P_{-\frac{3}{2}}(\frac{\pi}{2},\frac{\pi}{2})]
   +\frac{1}{3}\bigg(\frac{1}{2}\bigg)^3[P_{\frac{1}{2}}(\frac{\pi}{2},\frac{\pi}{2}) -P_{-\frac{1}{2}}(\frac{\pi}{2},\frac{\pi}{2})] \nonumber\\
   &-\frac{5}{12}\bigg(\frac{3}{2}\bigg)[P_{\frac{3}{2}}(\frac{\pi}{2},\frac{\pi}{2}) -P_{-\frac{3}{2}}(\frac{\pi}{2},\frac{\pi}{2})] -\frac{5}{12}\bigg(\frac{1}{2}\bigg)[P_{\frac{1}{2}}(\frac{\pi}{2},\frac{\pi}{2}) -P_{-\frac{1}{2}}(\frac{\pi}{2},\frac{\pi}{2})],\label{f:meansigmaxxy}
 \end{align}
where $\phi_x= \arctan \frac{1}{2} $. Applying the exchange symmetry yields the similar results for $\Sigma^{yzz}$, $\Sigma^{zzx}$ and $\Sigma^{zxx}$ if the the polar and azimuthal angles are chosen properly.

\begin{align}
 \langle\Sigma^{xxx}\rangle
 &=\bigg(\frac{3}{2}\bigg)^3[P_{\frac{3}{2}}(\frac{\pi}{2},0)-P_{-\frac{3}{2}}(\frac{\pi}{2},0)] +\bigg(\frac{1}{2}\bigg)^3[P_{\frac{1}{2}}(\frac{\pi}{2},0)-P_{-\frac{1}{2}}(\frac{\pi}{2},0)] \nonumber\\
 &-\frac{15}{8}[P_{\frac{3}{2}}(\frac{\pi}{2},0)-P_{-\frac{3}{2}}(\frac{\pi}{2},0)] -\frac{5}{8}[P_{\frac{1}{2}}(\frac{\pi}{2},0)-P_{-\frac{1}{2}}(\frac{\pi}{2},0)]. \label{f:meansigmaxxx}
\end{align}
One can obtain the similar results for $\Sigma^{zzz}$ by changing $x$ to $z$. For completeness, we also calculate the other polarizations.
\begin{align}
  &\langle \Sigma^x \rangle = \frac{3}{2}[P_{\frac{3}{2}}{(\frac{\pi}{2},0)}-P_{-\frac{3}{2}}{(\frac{\pi}{2},0)}] +\frac{1}{2}[P_{\frac{1}{2}}{(\frac{\pi}{2},0)}-P_{-\frac{1}{2}}{(\frac{\pi}{2},0)}], \label{f:meansigmax} \\
  &\langle \Sigma^{zz} \rangle = \frac{9}{4}[P_{\frac{3}{2}}{(0,0)}+P_{-\frac{3}{2}}{(0,0)}] +\frac{1}{4}[P_{\frac{1}{2}}{(0,0)}+P_{-\frac{1}{2}}{(0,0)}]-\frac{5}{4}, \label{f:meansigmazz} \\
  &\langle \Sigma^{xy} \rangle = \frac{9}{4}[P_{\frac{3}{2}}{(\frac{\pi}{2},\frac{\pi}{4})} +P_{-\frac{3}{2}}{(\frac{\pi}{2},\frac{\pi}{4})}] +\frac{1}{4}[P_{\frac{1}{2}}{(\frac{\pi}{2},\frac{\pi}{4})} +P_{-\frac{1}{2}}{(\frac{\pi}{2},\frac{\pi}{4})}]. \label{f:meansigmaxy}
\end{align}
The spin vector polarizations $\Sigma^y$ and $\Sigma^z$ have the similar results to $\Sigma^x$. They can be obtained by replacing the polar and azimuthal angles properly. The spin tensor polarizations $\Sigma^{xx}$, $\Sigma^{xy}$ and $\Sigma^{yz}$ can also be calculated in the same way.
We show the other polarizations in the Appendix.

\section{Applications} \label{S:application}

In the previous sections, we obtain 7 spin polarization components for spin-3/2 particles, $S_{LLL}$, $S_{LLT}^x$, $S_{LLT}^y$, $S_{LTT}^{xx}$, $S_{LTT}^{xy}$, $S_{TTT}^{xxx}$ and $S_{TTT}^{xxy}$, which correspond to Eqs. (\ref{f:sigmaSxyz})-(\ref{f:sigmaSzzz}). Here we note that one can freely choose any combination of the 7 polarizations as long as $S_{LTT}^{xy}$ is included. With these polarizations, it is straightforward to apply them into high energy reactions, e.g, baryon production electron-positron annihilation process. To illustrate the applications, we first give the following definitions.
\begin{align}
  & |S_{LLT}| =\sqrt{(S_{LLT}^x)^2+(S_{LLT}^y)^2}, \\
  & |S_{LTT}^x| =\sqrt{(S_{LTT}^{xx})^2+(S_{LTT}^{xy})^2}, \\
  & |S_{TTT}^{xx}| =\sqrt{(S_{TTT}^{xxx})^2+(S_{TTT}^{xxy})^2}, \\
  & S_{LLT}^x= |S_{LLT}| \cos\phi_{LLT}, \label{f:SLLTx}\\
  & S_{LLT}^y= |S_{LLT}| \sin\phi_{LLT},  \label{f:SLLTy}\\
  & S_{LTT}^{xx}= |S_{LTT}^x| \cos\phi_{LTT}, \\
  & S_{LTT}^{xy}= |S_{LTT}^x| \sin\phi_{LTT}, \\
  & S_{TTT}^{xxx}= |S_{LTT}^{xx}| \cos\phi_{TTT}, \label{f:STTTxxx}\\
  & S_{TTT}^{xxy}= |S_{LTT}^{xx}| \sin\phi_{TTT}. \label{f:STTTxxy}
\end{align}
We note that $S_{LTT}^{K_1}$ and $S_{TTT}^{K_2}$ ($K_1=xx, xy; K_2=xxx, xxy $) are defined respectively by $|S_{LTT}^{x}|$ and $|S_{TTT}^{xx}|$ rather than $|S_{LTT}|$ and $|S_{TTT}|$ because of the convenient measurements in experiments. To see this, we consider the inclusive spin-3/2 particle production process in electron-positron annihilation reaction.

It is known that the cross section of the annihilation process can be written as the product of the leptonic tensor and the hadronic tensor \cite{Chen:2016moq,Yang:2017sxz,Wei:2013csa,Wei:2014pma}. The hadronic tensor can not be calculated perturbatively because it contains the nonperturbative hadronization process. From the theoretical point of view, there are two kinds of method to calculate the hadronic tensor. The first one is to use the parton model~\cite{Feynman:1969ej,Feynman:1973xc,Bjorken:1969ja} which ensures that the hadronic tensor can be decomposed with twist \cite{Peskin:1995ev}. One is to decompose the hadronic tensor with the basic Lorentz tensors as shown in ref.~\cite{Chen:2016moq,Pitonyak:2013dsu}. In this paper, we only consider the second one.

For a general consideration, the hadronic tensor can be decomposed as
\begin{align}
   W^{S\mu\nu} &=\sum_{\sigma,j} W_{\sigma j}^S h_{\sigma j}^{S\mu\nu} + \sum_{\sigma, j} \tilde W_{\sigma j}^S \tilde h_{\sigma j}^{S\mu\nu},\label{f:Wsmunu}\\
   W^{A\mu\nu} &=\sum_{\sigma,j} W_{\sigma j}^A h_{\sigma j}^{A\mu\nu} + \sum_{\sigma, j} \tilde W_{\sigma j}^A \tilde h_{\sigma j}^{A\mu\nu},\label{f:Wamunu}
\end{align}
where $h_{\sigma j}^{S/A\mu\nu}$ are named as basic Lorentz tensor. We use $\thicksim$ to denote the parity violating terms (for BESIII energy scale, parity violating terms which come from the weak interaction vanish), $W_{\sigma j}^{S/A}$ are coefficients, subscript~$\sigma$ and $j$ denote the polarization and the serial number of the basic Lorentz tensor. To construct the basic Lorentz tensor, only the momenta of the virtual photon ($q$) and the produced hadron ($p$) can be used for the unpolarized case, i.e,
\begin{align}
& h_{Ui}^{S\mu\nu} = \Bigl\{ g^{\mu\nu}-\frac{q^\mu q^\nu}{q^2}, ~~ p_{q}^{\mu} p_{q}^{\nu}  \Bigr\}, \label{f:hUS}\\
& \tilde h_{U}^{A\mu\nu} = \varepsilon^{\mu\nu qp},\label{f:hUA}
\end{align}
where $g^{\mu\nu}$ is metric tensor, $p_{q}^\mu=p^\mu-q^\mu(p\cdot q)/q^2$. $p_q$ is used to keep the hadronic tensor satisfying the current conservation, $q_\mu W^{\mu\nu}=q_\nu W^{\mu\nu}=0$. For the spin vector terms, we have
\begin{align}
& h_{V}^{S\mu\nu} = \varepsilon^{\{\mu q p S} p_{q}^{\nu\}}, \\
& \tilde h_{Vi}^{S\mu\nu} = \Bigl\{ (q \cdot S) h_{Uj}^{S\mu\nu}, ~~S_q^{\{\mu} p_{q}^{\nu\}} \Bigr\}, \label{eq:thVS} \\
& h_{Vi}^{A\mu\nu} = \Bigl\{ (q \cdot S) \tilde h_{U}^{A\mu\nu}, ~~\varepsilon^{[\mu q p S} p_{q}^{\nu]} \Bigr\}, \\
& \tilde h_{V}^{A\mu\nu} = S_q^{[\mu} p_{q}^{\nu]},
\end{align}
where $S$ denotes the spin polarization vector $(S_L, S_T^x, S_T^y)$. The conventions $A^{[\mu}B^{\nu]} \equiv A^\mu B^\nu -A^\nu B^\mu$ and $A^{\{\mu}B^{\nu\}} \equiv A^\mu B^\nu +A^\nu B^\mu$ are also used. For the spin tensor case, it is convenient to use the following definitions,
\begin{align}
& h_{LLi}^{S\mu\nu} =S_{LL} h_{Ui}^{S\mu\nu}, \quad \quad  \tilde h_{LL}^{A\mu\nu} = S_{LL} \tilde h_{U}^{A\mu\nu}, \label{f:hLLA}\\
& h_{LT}^{S\mu\nu} = S_{LT}^{\{\mu} p_{q}^{\nu\}}, \quad \quad \ \  \tilde h_{LT}^{S\mu\nu} = \varepsilon^{\{\mu q p S_{LT}} p_{q}^{\nu\}}, \label{f:hLTtS}\\
& h_{LT}^{A\mu\nu} = S_{LT}^{[\mu} p_{q}^{\nu]},\quad \quad \ \  \tilde h_{LT}^{A\mu\nu} = \varepsilon^{[\mu q p S_{LT}} p_{q}^{\nu]}, \label{f:hLTtA}\\
& h_{TTi}^{S\mu\nu} = S_{TT}^{\mu\nu},  \quad \quad \quad \ \  \tilde h_{TT}^{S\mu\nu} = \varepsilon^{\{\mu \alpha q p} S_{TT\alpha}^{\nu\}}. \label{f:hTTtS}
\end{align}

From the above discussion, the construction of the basic Lorentz tensor can be straightforwardly extended to the high order tensor case. These basic Lorentz tensor which are related to $S_{LLL}$ can be written as
\begin{align}
& h_{LLLi,in}^{S\mu\nu} =S_{LLL} h_{Ui}^{S\mu\nu},  \quad \quad  \tilde h_{LLL,in}^{A\mu\nu} = S_{LLL} \tilde h_{U}^{A\mu\nu}.
\end{align}
The other basic Lorentz tensors can be constructed according to Eqs. (\ref{f:SLLTx})-(\ref{f:STTTxxy}) and Eqs. (\ref{f:hLTtS})-(\ref{f:hLTtA}) by replacing $S_{LT}$ to the corresponding polarization components, i.e,
\begin{align}
  & \tilde h_{LLT}^{S\mu\nu} = S_{LLT}^{\{\mu} p_{q}^{\nu\}},  \quad \quad   h_{LLT}^{S\mu\nu} = \varepsilon^{\{\mu q p S_{LLT}} p_{q}^{\nu\}}, \label{f:hLLTtS}\\
  & \tilde h_{LLT}^{A\mu\nu} = S_{LLT}^{[\mu} p_{q}^{\nu]}, \quad \quad   h_{LLT}^{A\mu\nu} = \varepsilon^{[\mu q p S_{LLT}} p_{q}^{\nu]}. \label{f:hLLTtA}
\end{align}
For $S_{LTT}^{x\mu}$ and $S_{TTT}^{xx\mu}$, they can be obtained in the same way, we do not repeat the calculation here.

Substituting these basic Lorentz tensors into Eqs. (\ref{f:Wsmunu})-(\ref{f:Wamunu}) and contracting with the leptonic tensor yields the cross section of the inclusive electron-positron annihilation process,
\begin{align}
  \frac{E_p d\sigma}{d^3 p}=\frac{\alpha_{em}^2}{s^2} \Bigl[
 \mathcal{F}_U +S_L \tilde{\mathcal{F}}_L+S_{LL}\mathcal{F}_{LL} + S_{LLL}\tilde{\mathcal{F}_{LLL}}\Bigr], \label{f:inclusive}
\end{align}
where $\alpha_{em}$ is the fine structure constant, $s=Q^2$. There are 15 polarized terms in the cross section. $\mathcal{F}_U$ denotes the unpolarized term. For simplicity, we only present the longitudinal polarized terms here. $\mathcal{F}_i$  are given by
\begin{align}
&{\cal F}_{U} =(1+\cos^2\theta) F_{U1}+  \sin^2\theta F_{U2} + \cos\theta F_{U3},\label{f:FU} \\
&\tilde{\cal F}_{L}=(1+\cos^2\theta)\tilde F_{L1}+\sin^2\theta \tilde F_{L2}+\cos\theta \tilde F_{L3}, \label{f:tFL} \\
&{\cal F}_{LL}= (1+\cos^2\theta) F_{LL 1}+\sin^2\theta F_{LL2}+\cos\theta F_{LL3}, \label{f:FLL} \\
&\tilde{\cal F}_{LLL}=(1+\cos^2\theta)\tilde F_{LLL1}+\sin^2\theta \tilde F_{LLL2}+\cos\theta \tilde F_{LLL3}, \label{f:tFLLL}
\end{align}
where $F_K (K=U1, \cdots, LLL3)$ are known as the structure functions which can be determined in experiments, $\theta$ is the scattering angel in the center of lepton momenta frame.

There are three steps to calculate the hadron polarizations according to the cross section. Since the main purpose of this article is to parameterize the spin-3/2 density matrix, we do not show the detailed calculations but show the major steps.  First of all, we calculate the probability of finding a certain spin operator (e.g, $S_{LTT}^{xy}=\langle \Sigma^{xyz}\rangle$) on its eigenstate $\psi_{LTT}$. We have done this in the previous section. Second, we need to calculate all eigenvalues of all the spin operators on eigenstate $\psi_{LTT}$. It is not difficult to solve the Schrodinger equations to calculate the eigenstate and these eigenvalues. There is another way to calculate these eigenvalues. Since we consider the polarizations on eigenstate $\psi_{LTT}$, in this case the SDM can be written as $\rho=|\psi_{LTT}\rangle \langle \psi_{LTT}|$. Using $|\psi_{LTT}\rangle \langle \psi_{LTT}|=$Eq. (\ref{f:rhoexpression}) yields all the eigenvalues. The final step is substituting all the eigenvalues on $|\psi_{LTT}\rangle$ into the cross section to obtain the terms which are only related to $S_{LTT}^{xy}$. Using Eq. (\ref{f:meansigmaxyz}), one can obtain the polarization.

To finish this section, we would like to stress again that the basic Lorentz tenors can be applied into the electron-positron annihilation, deeply inelastic scattering and the Drell-Yan processes. One can only change the four-momentum in accord with the process under considered.

\section{Summary} \label{S:summary}

The description of particles with spin can be attained by using a spin density matrix in high energy reaction. The spin density matrix can be written as a standard form by using the spherical tensor operator or a Cartesian form by using the irreducible spin tensor. In this paper we present a parametrization of the spin density matrix for spin-3/2 particles in the Cartesian form. We obtain four equations for the spin polarizations by comparing the standard form with the Cartesian form. These four equations can be used to eliminate the dependent spin polarizations in describing polarized system. We also present the probabilistic interpretations of these spin polarizations by defining the spin operator in the spherical coordinate frame. From Eq. (\ref{f:inclusive}), we can see that these results can be used in the cross section to extract hadron polarizations.

The SDM considered in this paper can not only be applied to the annihilation process but also applied to any other spin-3/2 baryon production processes. We obtain 7 spin polarization components for spin-3/2 particles, $S_{LLL}$, $S_{LLT}^x$, $S_{LLT}^y$, $S_{LTT}^{xx}$, $S_{LTT}^{xy}$, $S_{TTT}^{xxx}$ and $S_{TTT}^{xxy}$. As mentioned, one can freely choose any combination of the 7 polarizations as long as $S_{LTT}^{xy}$ is included. For spin-3/2 particle, $\Omega$ which is spin-3/2 is made up of three strange quarks. Measuring the polarization of it in the high energy reactions is an ideal way to study the strange quark spin transfer in hadronization processes, which can help us to understand more about the intrinsic structures of hadronization.

\section*{Acknowledgements}
{The authors thank Yukun Song and Kaibao Chen very much for their helpful discussions and suggestions. This work was supported by the National Laboratory Foundation (Grant No. 6142004180203).
}

\begin{appendix}

\section{Probabilistic interpretations of spin polarizations}

In Sect. \ref{S:probability}, we calculate part of the polarizations in terms of the probabilities. In this section, we show all the others. First of all, we show the spin vector polarizations.
\begin{align}
   &\langle \Sigma^y \rangle = \frac{3}{2}[P_{\frac{3}{2}}{(\frac{\pi}{2},\frac{\pi}{2})}-P_{-\frac{3}{2}}{(\frac{\pi}{2},\frac{\pi}{2})}] +\frac{1}{2}[P_{\frac{1}{2}}{(\frac{\pi}{2},\frac{\pi}{2})}-P_{-\frac{1}{2}}{(\frac{\pi}{2},\frac{\pi}{2})}], \\
   &\langle \Sigma^z \rangle = \frac{3}{2}[P_{\frac{3}{2}}{(0,0)}-P_{-\frac{3}{2}}{(0,0)}] +\frac{1}{2}[P_{\frac{1}{2}}{(0,0)}-P_{-\frac{1}{2}}{(0,0)}].
\end{align}

The spin vector polarizations are expressed as follows.
\begin{align}
  &\langle \Sigma^{xx} \rangle = \frac{9}{4}[P_{\frac{3}{2}}{(\frac{\pi}{2},0)}+P_{-\frac{3}{2}}{(\frac{\pi}{2},0)}] +\frac{1}{4}[P_{\frac{1}{2}}{(\frac{\pi}{2},0)}+P_{-\frac{1}{2}}{(\frac{\pi}{2},0)}]-\frac{5}{4},  \\
  &\langle \Sigma^{xz} \rangle = \frac{9}{4}[P_{\frac{3}{2}}{(\frac{\pi}{4},0)} +P_{-\frac{3}{2}}{(\frac{\pi}{4},0)}] +\frac{1}{4}[P_{\frac{1}{2}}{(\frac{\pi}{4},0)} +P_{-\frac{1}{2}}{(\frac{\pi}{4},0)}], \\
  &\langle \Sigma^{yz} \rangle = \frac{9}{4}[P_{\frac{3}{2}}{(\frac{\pi}{4},\frac{\pi}{2})} +P_{-\frac{3}{2}}{(\frac{\pi}{4},\frac{\pi}{2})}] +\frac{1}{4}[P_{\frac{1}{2}}{(\frac{\pi}{4},\frac{\pi}{2})} +P_{-\frac{1}{2}}{(\frac{\pi}{4},\frac{\pi}{2})}].
\end{align}

\begin{align}
    \langle\Sigma^{zzz}\rangle
    &=\bigg(\frac{3}{2}\bigg)^3[P_{\frac{3}{2}}(0,0)-P_{-\frac{3}{2}}(0,0)] +\bigg(\frac{1}{2}\bigg)^3[P_{\frac{1}{2}}(0,0)-P_{-\frac{1}{2}}(0,0)] \nonumber\\
    &-\frac{15}{8}[P_{\frac{3}{2}}{(0,0)}-P_{-\frac{3}{2}}{(0,0)}] -\frac{5}{8}[P_{\frac{1}{2}}{(0,0)}-P_{-\frac{1}{2}}{(0,0)}], \\
   \langle\Sigma^{yzz}\rangle
   &=\frac{5\sqrt{5}}{3}\bigg(\frac{3}{2}\bigg)^3[P_{\frac{3}{2}}(\theta_y,\frac{\pi}{2}) -P_{-\frac{3}{2}}(\theta_y,\frac{\pi}{2})] +\frac{5\sqrt{5}}{3}\bigg(\frac{1}{2}\bigg)^3[P_{\frac{1}{2}}(\theta_y,\frac{\pi}{2}) -P_{-\frac{1}{2}}(\theta_y,\frac{\pi}{2})] \nonumber\\
   &-\frac{4\sqrt{2}}{3}\bigg(\frac{3}{2}\bigg)^3[P_{\frac{3}{2}}(\frac{\pi}{4},\frac{\pi}{2}) -P_{-\frac{3}{2}}(\frac{\pi}{4},\frac{\pi}{2})] -\frac{4\sqrt{2}}{3}\bigg(\frac{1}{2}\bigg)^3[P_{\frac{1}{2}}(\frac{\pi}{4},\frac{\pi}{2}) -P_{-\frac{1}{2}}(\frac{\pi}{4},\frac{\pi}{2})] \nonumber\\
   &-2\bigg(\frac{3}{2}\bigg)^3[P_{\frac{3}{2}}(0,0) -P_{-\frac{3}{2}}(0,0)] -2\bigg(\frac{1}{2}\bigg)^3[P_{\frac{1}{2}}(0,0) -P_{-\frac{1}{2}}(0,0)] \nonumber\\
   &+\frac{1}{3}\bigg(\frac{3}{2}\bigg)^3[P_{\frac{3}{2}}(\frac{\pi}{2},\frac{\pi}{2}) -P_{-\frac{3}{2}}(\frac{\pi}{2},\frac{\pi}{2})]
   +\frac{1}{3}\bigg(\frac{1}{2}\bigg)^3[P_{\frac{1}{2}}(\frac{\pi}{2},\frac{\pi}{2}) -P_{-\frac{1}{2}}(\frac{\pi}{2},\frac{\pi}{2})] \nonumber\\
   &-\frac{5}{12}\bigg(\frac{3}{2}\bigg)[P_{\frac{3}{2}}(\frac{\pi}{2},\frac{\pi}{2}) -P_{-\frac{3}{2}}(\frac{\pi}{2},\frac{\pi}{2})] -\frac{5}{12}\bigg(\frac{1}{2}\bigg)[P_{\frac{1}{2}}(\frac{\pi}{2},\frac{\pi}{2}) -P_{-\frac{1}{2}}(\frac{\pi}{2},\frac{\pi}{2})], \\
   \langle\Sigma^{zzx}\rangle
   &=\frac{5\sqrt{5}}{3}\bigg(\frac{3}{2}\bigg)^3[P_{\frac{3}{2}}(\theta_{zx},0) -P_{-\frac{3}{2}}(\theta_{zx},0)] +\frac{5\sqrt{5}}{3}\bigg(\frac{1}{2}\bigg)^3[P_{\frac{1}{2}}(\theta_{zx},0) -P_{-\frac{1}{2}}(\theta_{zx},0)] \nonumber\\
   &-\frac{4\sqrt{2}}{3}\bigg(\frac{3}{2}\bigg)^3[P_{\frac{3}{2}}(\frac{\pi}{4},0) -P_{-\frac{3}{2}}(\frac{\pi}{4},0)] -\frac{4\sqrt{2}}{3}\bigg(\frac{1}{2}\bigg)^3[P_{\frac{1}{2}}(\frac{\pi}{4},0) -P_{-\frac{1}{2}}(\frac{\pi}{4},0)] \nonumber\\
   &-2\bigg(\frac{3}{2}\bigg)^3[P_{\frac{3}{2}}(0,0) -P_{-\frac{3}{2}}(0,0)] -2\bigg(\frac{1}{2}\bigg)^3[P_{\frac{1}{2}}(0,0) -P_{-\frac{1}{2}}(0,0)] \nonumber\\
   &+\frac{1}{3}\bigg(\frac{3}{2}\bigg)^3[P_{\frac{3}{2}}(\frac{\pi}{2},0) -P_{-\frac{3}{2}}(\frac{\pi}{2},0)]
   +\frac{1}{3}\bigg(\frac{1}{2}\bigg)^3[P_{\frac{1}{2}}(\frac{\pi}{2},) -P_{-\frac{1}{2}}(\frac{\pi}{2},0)] \nonumber\\
   &-\frac{5}{12}\bigg(\frac{3}{2}\bigg)[P_{\frac{3}{2}}(\frac{\pi}{2},0)-P_{-\frac{3}{2}}(\frac{\pi}{2},0)] -\frac{5}{12}\bigg(\frac{1}{2}\bigg)[P_{\frac{1}{2}}(\frac{\pi}{2},0) -P_{-\frac{1}{2}}(\frac{\pi}{2},0)], \\
   \langle\Sigma^{zxx}\rangle
   &=\frac{5\sqrt{5}}{3}\bigg(\frac{3}{2}\bigg)^3[P_{\frac{3}{2}}(\theta_{xz},0) -P_{-\frac{3}{2}}(\theta_{xz},0)] +\frac{5\sqrt{5}}{3}\bigg(\frac{1}{2}\bigg)^3[P_{\frac{1}{2}}(\theta_{xz},0) -P_{-\frac{1}{2}}(\theta_{xz},0)] \nonumber\\
   &-\frac{4\sqrt{2}}{3}\bigg(\frac{3}{2}\bigg)^3[P_{\frac{3}{2}}(\frac{\pi}{4},0) -P_{-\frac{3}{2}}(\frac{\pi}{4},0)] -\frac{4\sqrt{2}}{3}\bigg(\frac{1}{2}\bigg)^3[P_{\frac{1}{2}}(\frac{\pi}{4},0) -P_{-\frac{1}{2}}(\frac{\pi}{4},0)] \nonumber\\
   &-2\bigg(\frac{3}{2}\bigg)^3[P_{\frac{3}{2}}(\frac{\pi}{2},0) -P_{-\frac{3}{2}}(\frac{\pi}{2},0)] -2\bigg(\frac{1}{2}\bigg)^3[P_{\frac{1}{2}}(\frac{\pi}{2},0) -P_{-\frac{1}{2}}(\frac{\pi}{2},0)] \nonumber\\
   &+\frac{1}{3}\bigg(\frac{3}{2}\bigg)^3[P_{\frac{3}{2}}(0,0) -P_{-\frac{3}{2}}(0,0)]
   +\frac{1}{3}\bigg(\frac{1}{2}\bigg)^3[P_{\frac{1}{2}}(0,) -P_{-\frac{1}{2}}(0,0)] \nonumber\\
   &-\frac{5}{12}\bigg(\frac{3}{2}\bigg)[P_{\frac{3}{2}}(0,0)-P_{-\frac{3}{2}}(0,0)] -\frac{5}{12}\bigg(\frac{1}{2}\bigg)[P_{\frac{1}{2}}(0,0) -P_{-\frac{1}{2}}(0,0)],
 \end{align}
where $\theta_y=\arctan{\frac{1}{2}}, \theta_{zx}=\arctan{\frac{1}{2}}, \theta_{xz}=\arctan 2$.

\end{appendix}

\end{document}